\documentclass{llncs}

\usepackage[T1]{fontenc}
\usepackage[utf8]{inputenc}
\usepackage{todonotes}
\usepackage{url}
\usepackage{graphicx}
\usepackage{multirow}
\usepackage{footnote}
\usepackage{tablefootnote}

\begin{document}
	
	\title{Analysis of Footnote Chasing and Citation Searching in an Academic Search Engine} 
    
	\titlerunning{Analysis of Footnote Chasing and Citation Searching}  
	%
	\author{Ameni Kacem and Philipp Mayr}
	\authorrunning{Kacem and Mayr} 
	\institute{GESIS -- Leibniz Institute for the Social Sciences, Cologne, Germany\\
		\email{[ameni.sahraoui,philipp.mayr]@gesis.org}
	}
	
	\maketitle              

	\begin{abstract}		
In interactive information retrieval, researchers consider the user behavior towards systems and search tasks in order to adapt search results by analyzing their past interactions. In this paper, we analyze the user behavior towards Marcia Bates' search stratagems such as 'footnote chasing' and 'citation search' in an academic search engine. We performed a preliminary analysis of their frequency and stage of use in the social sciences search engine sowiport. In addition, we explored the impact of these stratagems on the whole search process performance. We can conclude that the appearance of these two search features in real retrieval sessions lead to an improvement of the precision in terms of positive interactions with 16\% when using footnote chasing and 17\% for the citation search stratagem.

     \end{abstract}
	

\section{Introduction}
\label{sec:intro}
Interactive information retrieval (IIR) refers to a research discipline that studies the interaction between the user and the search system. In fact, researchers have moved from considering only the current query and results set to focus more on the user's past interactions and the analysis of whole retrieval sessions. Research approaches aim to understand the user searching behavior in order to improve the ranking of results after submitting a query and enhance the user experience within an IR system.

In Digital Libraries (DLs), researchers study concepts such as search strategies \cite{Carevic:2016}, search term suggestions \cite{Hienert:2016,Mayr2016BIRNDL}, communities' detection \cite{Akbar:2012}, personalization of search results \cite{Liu2012}, recommendation's impact \cite{Hienert:2016}, user’s information needs frequency and change. In addition, many interactive IR models have been proposed in the literature (e.g. \cite{Ellis:1989}) that describe the user’s behavior by different steps (stages) of information seeking and interacting with an IR system. 

Similarly, in the academic search engine sowiport\footnote{http://sowiport.gesis.org/} \cite{Hienert2015}, we aim at understanding the user behavior in order to support him/her during the search session. In fact, DL users behave differently when interacting with the system as underlined by Bates \cite{Bates} who highlighted different concepts such as moves, tactics, stratagems, and strategies.

The goal of this paper is to explore the specific stratagems "footnote chasing" and "citation searching" which are often utilized as exploratory search functionalities in DLs \cite{Carevic:2016}. According to Bates \cite{Bates} \textit{footnote chasing} is defined as checking the reference and related material of a work backward in time. \textit{Citation searching} refers to a forward chaining of works citing another one through a citation index. Both stratagems are important search features which are build in state-of-the-art academic search engine like Google Scholar or ACM Digital Library and support the natural search behavior of a majority of academic searchers. 

Stratagems in general are not always supported by DLs because most of the search functions available in academic search engines remain on the "moves" or "tactics" level (described in \cite{Bates}) or cited references are completely missing in the system.
However, the use of stratagems can enhance the search experience of the user and this represents the focus of this work. 
In particular, we address the following research questions:\\ 
\textbf{RQ 1: Which usage patterns can be observed from footnote chasing (FC) and citation searching (CS)?}

In this paper, we analyze the usage patterns of footnote chasing (FC) and citation searching (CS) stratagems in real retrieval sessions in terms of frequency of their use and the stage at which they appear.\\ 
\textbf{RQ 2: How successful are retrieval sessions using the FC and CS stratagems?}

The use of a stratagem can impact the session conduct in different ways. We examine the interactions of the users in sowiport DL in order to measure the usefulness and the precision of sessions having such stratagems. We determine the session success based on the presence of positive actions proposed recently by Hienert and Mutschke \cite{Hienert:2016}. In the following, we measure the amount of positive actions before and after our two stratagems' (FC and CS) occur in a retrieval session. 

The remainder of this paper is organized as follows. In the next section, we present an short overview of basic concepts explored by researchers in the field of DLs. In Section ~\ref{sec:bibliometric} we analyze the user behavior towards the FC and CS stratagems and how using them affects the quality of the whole session search. Finally, we summarize our findings and present some perspectives relevant for future work. 

\section{Related Work}  
\label{sec:related}
Bates \cite{Bates} has specified different types of user behavior toward search system, among them we cite: \textit{moves}, \textit{tactics}, \textit{stratagems} and \textit{strategies}. A move refers to a basic action performed by the user. A tactic resides in using additional moves to go with the search. As for stratagems, they indicate complex and multiple moves/tactics having knowledge of a particular search domain. A strategy is a combination of moves, tactics and stratagems as a plan to pursue during the search session.
Footnote chasing and citation searching are popular \textit{stratagems} 
that refers to the study of documents and their bibliographic references and citations. 
Schneider and Borlund \cite{Schneider:2004} studied the effectiveness of using stratagems in constructing and maintaining thesauri vocabulary and structure.
Mahoui and Cunningham \cite{mahoui:2001} specified the importance of understanding the information of DL users in creating useful and stable search systems. 
They analyzed transaction logs to study usage patterns of CiteSeer in terms of query and search patterns.
Xie \cite{Xie2002} analyzed the users' search behaviors and their relationships with their information needs by specifying a hierarchical level of users' goals.
Shute and Smith \cite{SHUTE1993} identified 13 knowledge-based tactics arranged into three categories: broaden topic scope, narrow topic scope and change topic scope.
Carevic and Mayr \cite{Carevic:2015} proposed bibliometric-enhanced search facilities such as "journal run" or "citation search" and their possible integration in DLs. 
In their position paper, they argue that bibliometric-enhanced stratagems can facilitate domain specific search activities by applying bibliometric measures for re-ranking and/or rearranging DL-entities like documents, journals or authors. They propose different types of stratagem implementations like "extended journal run" or "context-preserving journal run" or extended versions of citation search.

\section{Methodology} 
\label{sec:bibliometric}
In this section, we first provide details about the data set that we used for our analysis. Then, we describe the approach used to answer research questions raised in Section~\ref{sec:intro}. 

\subsection{Data Set}
\label{sec:dataset}
Sowiport\footnote{http://www.sowiport.de} is a DL for the Social Sciences that contains more than nine million records, full texts and research projects included from twenty-two different databases whose content is in English and German \cite{Hienert2015}. For a part of the collections, namely the ProQuest databases "Sociological Abstracts", "Social Services Abstracts", "Applied Social Sciences Index and Abstracts", "Worldwide Political Science Abstracts" and "Physical Education Index", sowiport provides references and builds a citation index over its collections. These references and citations are part of the analysis in the following.

The \textit{Sowiport User Search Sessions Data Set (SUSS)}\footnote{http://dx.doi.org/10.7802/1380 and \cite{Mayr2016}} contains individual search sessions extracted from the transaction log of sowiport. The data was collected over a period of one year (between 2nd April 2014 and 2nd April 2015)\footnote{A detailed description of the data set can be found in \cite{Mayr2017}.}. The web server log files and specific JavaScript-based logging techniques were used to capture the user behavior within the system. The log was heavily filtered to exclude transactions performed by robots. All transaction activities are mapped to a list of 58 different user actions which cover all types of activities and pages that can be carried out/visited within the system (e.g. typing a query, visiting a document, selecting a facet, exporting a document, etc.). For each action, a session id, the date stamp and additional information (e.g. query terms, document ids, and result lists) are stored. Based on the session id and date stamp, the step in which an action is conducted and the length of the action is included in the data set as well. The session id is assigned via browser cookies and allows tracking user behavior over multiple searches. Thus, in the data set we find 484,449 individual search sessions and a total of 7,982,427 log entries.

\subsection{Description of Actions in the Session Log}
Searching sowiport can be performed through an \textit{All fields} search box (default search without specification), or through specifying one or more field(s): title, person, institution, number, keyword or year.
The users' main actions are described in Table~\ref{tab:actions}. In fact, we grouped the main actions into two categories: "Query"-related and "Document"-related actions. Another categorization of actions was proposed in \cite{Hienert:2016} by specifying search interactions and successive positive actions. 

\begin{table}
\centering
\caption{Main actions performed by users in sowiport}
\label{tab:actions}
\begin{tabular}{|l|l|p{8cm}|}
\hline
Category & Action & Description \\
\hline
 \multirow{8}{*}{Query} & query\_form & Formulating a query \\
    & search & A search result list for any kind of search \\
    & search\_advanced  &  A search with the advanced settings that can limit the search fields, information type, provider/database, language: or time (year, recent only) \\
    & search\_keyword & A search for a keyword  \\
    & search\_thesaurus & Usage of the thesaurus system\\
    & search\_institution & A search for an institution \\
    & search\_person & A search for a specific person (author/editor) \\
\hline 
\multirow{8}{*} {Document} & view\_record & Displaying a record in the result list after clicking on it\\
    & view\_citation & View the document's citation(s) \\
    & view\_references & View the document's references \\
    & view\_description & View the document's abstract \\
    & export\_bib & Export the document through different formats \\
    & export\_cite & Export the document's citations list \\
    & export\_mail & Send the document via email \\
    & to\_favorites & Save the document to the favorite list \\
\hline
\end{tabular}
\end{table}

Main user actions as described before can be categorized into actions regarding either search queries or documents. These actions are used in different scales in the data set. Query-related actions represent 29.84\% while document-related actions represent 35.79\% of the total amount of actions. The rest of actions contain navigational interactions such as logging in the system, managing favorites, and accessing the system pages.
 
In Table~\ref{tab:SessionSample}, we show a specific session, the user's ID and the actions' label and length in seconds. 
In this session, the user with ID \textit{41821} started with logging into the system and then submitted a query describing his/her information need (\textit{query\_form}) after doing some navigational actions. After getting the result list, labeled as \textit{resultlistids}, the user performed additional searches (\textit{searchterm\_2}), and displayed some results' content (\textit{view\_record}). Finally, he/she checked the external availability of a result (\textit{goto\_google\_scholar}). We notice that the user spent more than 40\% of the time reading documents' content.

\begin{table}
\centering
\caption{Sample of a session search for a specific user}
\label{tab:SessionSample}
\begin{tabular}{|c|p{3cm}|p{3.5cm}|l|}
\hline
User ID & Date & Action label & Action length (s)\\
\hline
\multirow{9}{*} {41821} & 2014-10-28 16:08:46 & goto\_login   & 1   \\
& 2014-10-28 16:08:47 & goto\_favorites       & 21  \\
& 2014-10-28 16:09:08 & goto\_home            & 2   \\
& 2014-10-28 16:09:13 & query\_form           & 22  \\
& 2014-10-28 16:09:35 & search                & 10  \\
 & 2014-10-28 16:09:35 & searchterm\_2         & 10  \\
& 2014-10-28 16:09:35 & resultlistids         & 10  \\
 & 2014-10-28 16:09:45 & view\_record          & 31  \\
& 2014-10-28 16:09:45 & docid                 & 31  \\
& 2014-10-28 16:10:16 & view\_record          & 392 \\
& 2014-10-28 16:17:07 & goto\_google\_scholar & 0  \\
\hline
\end{tabular}
\end{table}

In this paper, we are interested in the stratagems \textit{view\_citation} aka CS and \textit{view\_references} aka FC that are found in 20,353 sessions of the SUSS dataset. In our data set, we have 1,520 sessions within a user performed FC and 18,833 sessions with a CS.

\subsection{Measurements}
To answer the first research question described in Section~\ref{sec:intro}, we analyze the sessions with the mentioned stratagems FC and CS. 

For a session $S$ during which a set of interactions $\left \{ I \right \}$ is performed by the user, we define:
\begin{itemize}
\item $Strat$ is a stratagem such as FC and CS,
\item $Pos$ is a positive interaction present in our data set among the following set
$\left \{ P \right \}$ described in \cite{Hienert:2016}:\\ {goto\_fulltext, goto\_google\_scholar, goto\_local\_availability, goto\_google\_books,  view\_description, export\_cite, export\_bib, export\_mail, to\_favorites, export\_search\_mail, save\_search, save\_search\_history, save\_to\_multiple\_favorites}.\\
\end{itemize}

To answer the second research question, we measure the precision of a stratagem before 
($Precision(Strat)_b$) and after ($Precision(Strat)_a$) so we verify if it has an influence on the conduct of a session. We verify if we can find more positive actions after using a stratagem comparing to the number of positive actions before its utilization. 

\begin{equation}
Precision(Strat)_b = \left (\frac{\left | Pos \in \left \{ P \right \} \right |}{\left | I \right |}  \right )_b 
\label{eq:before}
\end{equation}

\begin{equation}
Precision(Strat)_a = \left (\frac{\left | Pos \in \left \{ P \right \} \right |}{\left | I \right |}  \right )_a
\label{eq:after}
\end{equation}

To have an overview of a stratagem benefit,  we measure the \textit{Usefulness} as the percentage of successful sessions in terms of positive actions among all the sessions including both of the studied stratagems. This measure is inspired from the \textit{Global Usefulness} measure proposed by \cite{Hienert:2016}:
\begin{equation}
Usefulness(Strat) = \frac{\left | s^{+}_{Strat} \right |}{\left | s_{Strat} \right |}
\label{eq:global}
\end{equation}
where $s^{+}(Strat)$ indicates session success in terms of positive actions occurrence after using a specific stratagem, and $\left | s_{Strat} \right |$ represents the number of sessions using a stratagem (footnote chasing or citation search) no matter the type of user's interactions (positive or not).

\section{Results} 
Our preliminary results show the distribution of stratagems at different stages of the sessions (see Figure 1 and Figure 2). 
We observe that citation search (Figure 1a) appear mostly in the end of sessions (90\%) with 5,400 sessions and in the middle (50\%) with 3,504 among 18,833 sessions presenting this stratagem. For footnote chasing (Figure 1b), it appears mostly in the middle of the session (50\% - 60\%) within 489 sessions among the 1,520 sessions including this stratagem. 

\begin{figure}[h]
\begin{center}
	\includegraphics[scale=0.5]{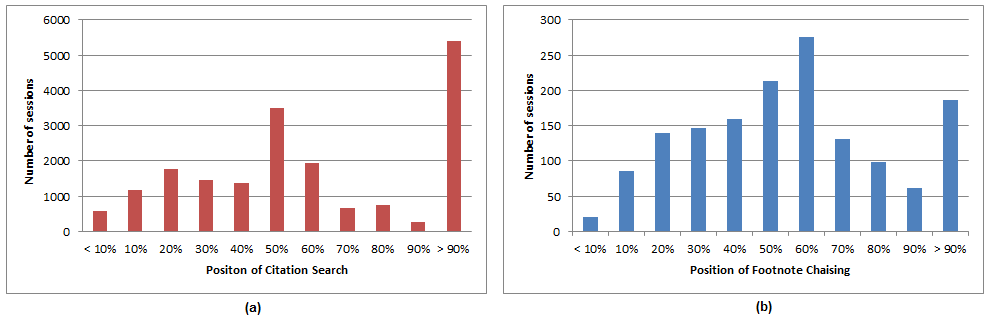}
	\caption{Number of sessions having stratagems and stages of their appearance in a session}
	\label{fig:Positions}
\end{center}
\end{figure}
\newpage

\begin{figure}[h]
\begin{center}
	\includegraphics[scale=0.45]{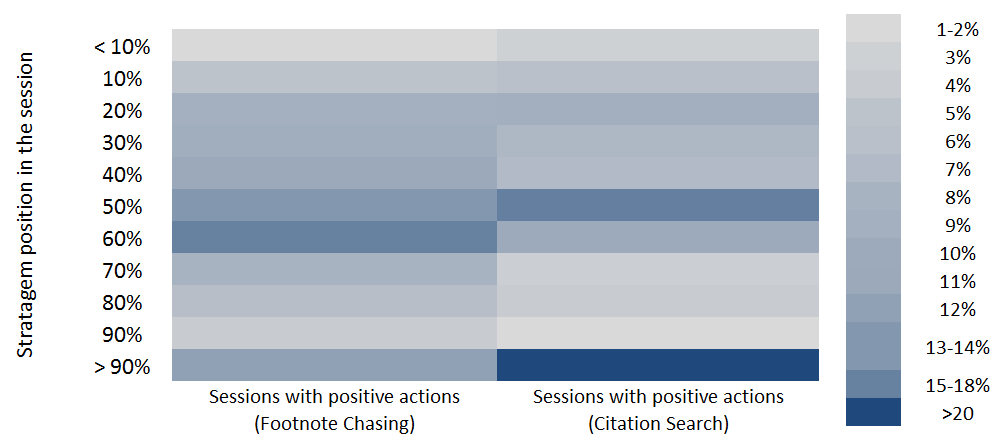} 
	\caption{Distribution of stratagems in the sessions (Footnote Chasing and Citation Search)}
	\label{fig:heatmap}
\end{center}
\end{figure}

We note that the position of the stratagem differs from one session to another due to the difference of sessions' length. We noticed from the user behavior analysis that for sessions that are short, the stratagem appears after or in the middle of sessions. For longer sessions, the stratagem appears between the first 30\% and 50\% of a session's interactions.

Then, in order to study the effectiveness of the stratagems \textit{Footnote chasing} and \textit{Citation searching}, we measure their precision before and after their appearance during search sessions. This measure is based on a set of positive actions that are considered as indicators of the session success \cite{Hienert:2016}.

In Table~\ref{tab:results}, we present the results of the measures described in equations~\ref{eq:before}, ~\ref{eq:after} and~\ref{eq:global}. We use as a baseline a set of random sessions that do not include both of the stratagems. 

\begin{table}[!htbp]
\centering
\caption{Evaluation of stratagems' use effect on the sessions' search}
\label{tab:results}
\begin{tabular}{|c|c|c|c|}
\hline
Measure & Footnote Chasing & Citation Search & Baseline\tablefootnote{The baseline is a set of 100 random sessions that do not include both of the stratagems.} \\
\hline
$Precision(Strat)_b$ & 0.0284 &  0.0197 & - \\
$Precision(Strat)_a$ & 0.1975 & 0.1897 & - \\
Gain In Precision\tablefootnote{Gain: computed as the difference between the precision-after and precision-before.}& 16.19\% & 17\%  & - \\ 
$Usefulness(Strat)$ & 77.2\% & 63.3\% & 19\% \\
\hline
\end{tabular}
\end{table}

From Table~\ref{tab:results}, we conclude that footnote chasing and citation search have a positive impact on the session performance in terms of positive actions appearance with 16.19\% and 17\% correspondingly. In fact, the positive actions appeared before a stratagem are lower in most of the cases to positive ones appeared after. We can present an improvement in terms of positive actions occurrence when a stratagem is employed by the users and thus conclude that these stratagems lead to successful sessions and positive interactions. In fact, there is a reformulation of the query in only 7.67\% of sessions with citation search and 12.63\% of those with footnote chasing. Thus, in most of the sessions, the positive actions are directly related to the first search and first appearance of stratagems.

As for the global performance of both stratagems, the usefulness values of 77.2\% for footnote chasing and 63.3\% for citation search are considered as promising results as the value of this measure is in $\left [ 0,1 \right ]$ and it improves the results of the baseline with over 44\%.

In Table~\ref{tab:posneg} we present the different ways in which a stratagem can effect the conduct of a session: positive, non-positive or neutral. We obtain those values through the difference $Diff(a,b)$ between the precision before ($Precision(Strat)_b$) and the one after ($Precision(Strat)_a$) the use of a stratagem.
We can see that both stratagems affect the sessions mostly in a positive way. 
This means that the use of a stratagem influences the user behavior and make him more interactive with the system in a beneficial aspect. Also, the non-positive effect is present in a small amount of sessions comparing to the neutral one. In fact, the absence of positive actions does not mean a negative conduct of a session because the user is always interacting with the system using moves and tactics that are not judged positive but not specified as negative either.

\begin{table}
\centering
\caption{Percentage of sessions with stratagems having three ways of effect on the sessions' conduct}
\label{tab:posneg}
\begin{tabular}{|c|c|c|}
\hline
 & Footnote Chasing & Citation Search \\
\hline
Positive effect ($Diff>0$) & 62.79\% & 56.54\% \\
Non-positive Effect ($Diff<0$)& 10.88\% & 6.77\% \\
Neutral Effect ($Diff=0$)& 26.33\% & 36.69\%  \\
\hline
\end{tabular}
\end{table}

\section{Summary}
In this paper, we started to investigate the use of two stratagems namely \textit{Footnote Chasing} and \textit{Citation Search} in sowiport digital library and more precisely in the SUSS data set \cite{Mayr2016}. In fact, studying the user behavior towards stratagems can enhance the user-system interactions and lead to more useful academic search engines \cite{Carevic:2015}. 
Using the SUSS data set, we examined the frequency and stage of use of such stratagems as well as their impact on sessions. We verify whether their utilization can lead to successful sessions. We measured the success of sessions by measuring the difference between the precision before and after using a stratagem. Both of the precisions are obtained thanks to the positive actions occurrence in sessions \cite{Hienert:2016}.

As future work, we need to perform further analysis of other stratagems such as journal run for instance and to go beyond log analysis to do user studies in order to compare user feedback with the findings of this study.
 

\section{Acknowledgement}
This work was funded by Deutsche Forschungsgemeinschaft (DFG), grant no. MA 3964/5-1; the AMUR project at GESIS together with the working group of Norbert Fuhr. 
The AMUR project aims at improving the support of interactive retrieval sessions following two major goals: improving user guidance and system tuning.

\bibliographystyle{splncs}
\bibliography{bibdb.bib}
\end{document}